\documentclass{sigchi}
\usepackage{siunitx}
\usepackage{float}
\usepackage{amsmath}
\usepackage[titlenumbered,ruled]{algorithm2e}
\usepackage{algpseudocode}
\usepackage{pifont}
\usepackage{times}
\usepackage{url}
\usepackage{tabulary}
\usepackage{booktabs}
\usepackage{multirow}
\usepackage{epstopdf}
\usepackage{tablefootnote}
\usepackage{footnote}
\usepackage{balance}  
\usepackage{graphics} 
\usepackage[T1]{fontenc}
\usepackage{txfonts}
\usepackage{mathptmx}
\usepackage[pdftex]{hyperref}
\usepackage{color}
\usepackage{textcomp}
\usepackage{subfig}
\usepackage{microtype} 
\usepackage{ccicons}  
\usepackage[utf8]{inputenc} 

\newcounter{NumberOfComments}
\stepcounter{NumberOfComments}

\newcounter{JSNumberOfComments}
\stepcounter{JSNumberOfComments}

\newcounter{JRNumberOfComments}
\stepcounter{JRNumberOfComments}

\def\plaintitle{}

\def\emptyauthor{}
\def\plainkeywords{}

\makeatletter
\def\url@leostyle{%
  \@ifundefined{selectfont}{
    \def\UrlFont{\sf}
  }{
    \def\UrlFont{\small\bf\ttfamily}
  }}
\makeatother
\def\pprw{8.5in}
\def\pprh{11in}

\definecolor{linkColor}{RGB}{6,125,233}
\hypersetup{%
  pdftitle={\plaintitle},
  pdfauthor={\emptyauthor},
  pdfkeywords={\plainkeywords},
  bookmarksnumbered,
  pdfstartview={FitH},
  colorlinks,
  citecolor=black,
  filecolor=black,
  linkcolor=black,
  urlcolor=linkColor,
  breaklinks=true,
}

\begin{document}

\title{White, Man, and Highly Followed: Gender and Race Inequalities in Twitter%
\titlenote{\textcolor{red}{\textbf{This is a pre-print of a paper accepted to appear at Web Intelligence 2017}}.\\ \\
\textcolor{blue}{White, Man, and Highly Followed: Gender and Race Inequalities in Twitter.\\
Johnnatan Messias, Pantelis Vikatos, and Fabrício Benevenuto\\
In Proceedings of the IEEE/WIC/ACM International Conference on Web Intelligence (WI'17). Leipzig, Germany. August 2017.\\\\
\textbf{Bibtex: \url{http://johnnatan.me/bibtex/messias_wi_2017.bib}\\
For more published work: \url{http://johnnatan.me}}}}}

\numberofauthors{3}
\author{%
\alignauthor
Johnnatan Messias\\
       \affaddr{Universidade Federal de Minas Gerais}\\
       \affaddr{Belo Horizonte, Brazil}\\
       \email{johnnatan@dcc.ufmg.br}
\alignauthor
Pantelis Vikatos\\
       \affaddr{University of Patras}\\
       \affaddr{Rio, Greece}\\
       \email{vikatos@ceid.upatras.gr}
\alignauthor
Fabricio Benevenuto\\
       \affaddr{Universidade Federal de Minas Gerais}\\
       \affaddr{Belo Horizonte, Brazil}\\
       \email{fabricio@dcc.ufmg.br}
}

\maketitle

\begin{abstract}
Social media is considered a democratic space in which people connect and interact with each other regardless of their gender, race, or any other demographic factor. Despite numerous efforts that explore demographic factors in social media, it is still unclear whether social media perpetuates old inequalities from the offline world. In this paper, we attempt to identify gender and race of Twitter users located in U.S. using advanced image processing algorithms from Face++. Then, we investigate how different demographic groups (i.e. male/female, Asian/Black/White) connect with other. We quantify to what extent one group follow and interact with each other and the extent to which these connections and interactions reflect in inequalities in Twitter.  Our analysis shows that users identified as White and male tend to attain higher positions in Twitter, in terms of the number of followers and number of times in user's lists.  We hope our effort can stimulate the development of new theories of demographic information in the online space.

\end{abstract}

\section{Introduction} \label{sec:intro}

Nowadays, millions of people constantly use online social networking sites, such as Facebook and Twitter, sharing content about their daily lives and things that happen around them.  These systems have revolutionized the way we communicate, by organizing our offline social relationships in a digital form.  

Increasingly, our society has been treating actions in the online and offline space in an indistinguishable way. It is now common to see cases in which content posted on Facebook or Twitter is used to vet job candidates, support divorce litigation, or terminate employees. Although these platforms provide a democrat space for conversations and debates, they also open space for unsolved issues of our society, such as gender and race inequalities. 

Indeed, inequalities and the main barriers that keep a given demographic from rising beyond a certain level in a hierarchy have been perceived and quantified in many online systems. For example, recent efforts show gender disparities on content production in Wikipedia~\cite{graells2015first,wagner2016women}. A recent study shows that Black tenants have much fewer chances of renting a place on Airbnb\footnote{\url{http://www.airbnb.com}}~\cite{edelman2017racial}. Also, some drivers for both UberX\footnote{\url{http://www.uber.com}} and Lyft\footnote{\url{http://www.lyft.com}} discriminate on the basis of the perceived race of the traveler~\cite{ge2016racial}. There is also evidence that users in Twitter perceived as female experience a `glass ceiling', similar to the barrier females face in attaining higher positions in companies~\cite{nilizadeh2016twitter}.  


Overall, identifying inequalities and asymmetries in demographics in the online world is crucial for the development of features that can promote equality and diversity in these systems.  However, despite the recent surge of interest in demographic aspects of users from social networks, inferring such demographic attributes have been usually challenging and limited due to the difficult to identify gather them from the users in these systems. 


Particularly, in Twitter, a number of efforts have explored gender differences by identifying gender based on the user's name \cite{blevins2015jane,karimi2016,nilizadeh2016twitter}. However, studies of race inequality in Twitter are still limited. In this paper, we focus on filling this gap by answering the following research question: \textit{Do Gender and Race affect online interactions and impact people's connections in Twitter?} To do that, we crawled a large scale sample of active Twitters and then we identify the gender and race of about $1.6$ million users located in U.S by using Face++\footnote{\url{http://www.faceplusplus.com}}, a face recognition software able to recognize gender and race of identifiable faces in the user's profile pictures. 

Our findings reinforce previous observations about disadvantages against female users in Twitter and also identify advantages for White users, in comparison with those identified as Black and Asian. Our analysis also characterizes how these different demographic groups interact among themselves, providing several insights about the perceived inequality causes. We believe our findings not only contribute to the social science literature, but we also hope that they can contribute to the design of link recommendation systems that can act to diminish such inequalities. As a final contribution, we plan to make our dataset available to the research community by the time of publication of this article. 

The rest of the paper is organized as follow. The Section~\ref{sec:relatedwork} briefly reviews related efforts in the literature. Then, in Section~\ref{sec:dataset}, we present the Twitter and demographic dataset. Next, Section~\ref{sec:results_inequalities} characterizes gender and race inequalities in visibility and in Section~\ref{sec:results_interconnections} in terms of connections and interactions in Twitter. Finally, Section~\ref{sec:conclusion} concludes the paper and offers directions of future work. 

\section{Related Work} \label{sec:relatedwork}
In this section, we review the related literature along two axes. First, we discuss the methodology used by efforts that measure demographic factors in Twitter. Then, we discuss the main efforts that have explored gender and race Inequality in social networks.  

\subsection{Demographics in Social Media} 

There are several studies that retrieve information from social media to analyze and predict real-world phenomena such as stock market~\cite{bollen2011twitter}, migration~\cite{Messias2016@asonam}, election~\cite{o2010tweets, tumasjan2010predicting}, and also political leaning of Twitter users \cite{juhi2017,pennacchiotti2011democrats}. 

In the field of demographics, one of the first efforts in demographics in Twitter compares geographic and demographic distribution of users of U.S. population \cite{mislove2011understanding}. After that, there have been many studies that extract demographic information, using different strategies and different purposes~\cite{blevins2015jane,karimi2016,liu2013s,nilizadeh2016twitter,reis2017,vikatos2017}. Particularly, Chen \textit{et al.}~\cite{chen2015comparative} focus on demographic inference using namely profile self-descriptions and profile images. They compare five signals in order to categorize users to their demographic status. They use users' names, self-descriptions, tweets, social networks, and profile images to infer demographic attributes (ethnicity, gender, age). \cite{culotta2015predicting} describe an alternative methodology assuming  that the demographic profiles  of visitors to a website are correlated with the demographic profiles of followers of that website on social network e.g. Twitter; and  propose a regression model  to declare demographic attributes (gender, age, ethnicity, education, income, and child status). Meanwhile, \cite{Nakandala@icwsm} show that female streamers receive significantly more objectifying comments while male streamers receive more game-related comments.

Another study, An \textit{et al.}~\cite{an2016greysanatomy} examine the correlation of hashtags that are used in different demographic groups, using Face++ and the user's profile pictures. They show that this strategy is reliable and provides accurate demographic information for gender and race. More recently, Chakraborty  \textit{et al.}~\cite{chakraborty2017@icwsm} analyze the demographic of Twitter Trending Topics, using Face++ as well. They use an extensive data collected from Twitter in order to make the first attempt to quantify and explore the demographic biases in the crowdsourced recommendations. They find that a large fraction of trends are promoted by crowds whose demographics are significantly different from the overall Twitter population. More worryingly, certain demographic groups are systematically under-represented among the promoters of the trending topics.

Our effort uses the similar strategy to gather demographic information, but we investigate very different research questions as we focus on inequalities and demographic aspects of users connections. We note that this strategy to gather demographic aspects, applied to active users within a country, allows us to take a step further in many existing studies on the field as it allows us to study race inequalities and also allows us to investigate how  connections are established among demographic groups (i.e. male/female, Asian/Black/White). We hope our new and large-scale dataset may largely contributes to this research field.

\subsection{Inequalities in Twitter Visibility}

Homophily states that similarity breeds social connections~\cite{mcpherson2001birds}. In our society, people tend to characterize and categorize others in terms of demographic aspects, such as gender and race~\cite{willis2006first}, making them intuitively natural concepts through which people might account similarities. The existing literature is still limited in explaining the inequalities that arise from the establishments of these social connections in Twitter. 

Most of the existing efforts in this space attempt to quantify and understand other factors that are highly correlated with visibility (i.e. number of followers). Particularly, Morales \textit{et al.}~\cite{morales2014efficiency} show that the reputation and the impact on the society of a user constitute a significant factor of the online visibility growth. Other efforts show that the attention that users gain is correlated with his/her online behavior in terms of the frequency of their interactions~\cite{Comarela:2012:UFA:2309996.2310017,romero2011influence,wu2011says}. Freitas \textit{et al.}~\cite{Freitas2015@asonam} created $120$ socialbots in Twitter, with different behaviors and demographics features, to show that users who post more, post about specific topics, reply, favorite or mention other frequently, may acquire more followers, and consequently, more visibility. 
Despite the extreme importance of these efforts, they do not investigate the correlation of demographic issues with visibility. Thus, our effort is complementary to theirs. 

There have been efforts quantifying differences and inequalities in many other social media, including Wikipedia~\cite{graells2015first,wagner2016women} and Pinterest~\cite{Gilbert:2013:INT:2470654.2481336}. Also, inequalities in  between conversation and objectification in different genders have been analyzed in online social gaming such as Twitch ~\cite{Nakandala@icwsm}.

However, the study that is closest related to ours explores gender inequalities in Twitter~\cite{nilizadeh2016twitter}. Authors show that gender may allow inequality to persist in terms of online visibility. Our paper further explores race as a new demographic dimension. More important, unlike previous studies, our work focuses on investigating how different demographic groups (i.e. male/female, Asian/Black/White) connect with other. We quantify to what extent one group follows and interacts with each other and the extent to which these connections and interactions reflect in inequalities in Twitter. 

\section{Demographic Information Dataset} \label{sec:dataset}

This section describes the data collection and the method to infer demographic information of individual Twitter users. Our ultimate goal consists of gathering a dataset containing active U.S. Twitter users, their demographic information, a sample of their connection graph (friends), and their tweets. Next, we describe our steps to create this dataset and also discuss its main limitations.

\subsection{Twitter Dataset Gathered}
To identify active Twitter users we gathered data from the $1\%$ random sample of all tweets, provided by the Twitter Stream API~\footnote{https://dev.twitter.com/streaming/public}. Our dataset covers a period of three complete months, from July to September 2016. In total, we collected $341,457,982$ tweets posted by around $50,270,310$ users.

We restricted our demographic studies to Twitter users located in the United States. As geographic coordinates are available on Twitter only for a limited number of users (i.e. $<$ $2\%$)~\cite{icwsm10cha}, our strategy to identify U.S. Twitter users was based on part of the methodology used in previous efforts~\cite{kulshrestha2012geographic,chakraborty2017@icwsm}. We have used the timezone information as a first filter and then we attempted to remove from this filtered dataset those users that provided free text location indicating they are not U.S. (i.e. Montreal, Vancouver, Canada). We end up with a dataset containing $6,286,477$ users likely located in the United States.  


\subsection{Crawling Demographic Information}


Most of the existing studies related to demographics of users in Twitter have looked into gender and age. 
Some efforts attempt to infer the user's gender from the user name~\cite{blevins2015jane,karimi2016,liu2013s,mislove2011understanding}. However, some users may not use proper names, consequently their gender could not be inferred properly~\cite{liu2013s}. Others have attempted to identify patterns like {\it `$25$ yr old'} or {\it `born in $1990$'} in Twitter profile description to identify the user age~\cite{sloan2015tweets}. 

Here, we use a different strategy that allows us to study another demographic dimension: the user's race. 
To do that, we crawl the profile picture Web link of all Twitter users identified as located within the United States. 
In December $2016$, we crawled the profile picture's URLs of about $6$ million users, discarding $4,317,834$ ($68.68\%$) of them.  We discarded users in two situations, first when the user does not have a profile picture and second when the user has changed her picture since our first crawl. When users change their picture, their profile picture URL changes as well, making it impossible for us to gather these users in a second crawl.   

From the remaining $1,968,643$ users, we submitted the profile picture Web links into the {\it Face++ API}. Face++ is a face recognition platform based on deep learning~\cite{fan2014learning,yin2015learning} able to identify the gender (i.e. male and female) and race (limited to Asian, Black, and White) from recognized faces in images. 

We have also discarded those users whose profile pictures do not have a recognizable face or have more than one recognizable face, according to Face++. Our baseline dataset, also used by Chakraborty \textit{et al.}~\cite{chakraborty2017@icwsm} work, contains $1,670,863$ users located in U.S. with identified demographic information. Table~\ref{table:expected} shows the demographic distribution of users in our dataset across the different demographic groups. We use this dataset to study inequalities in visibility that will be present in Section~\ref{sec:results_inequalities}. 

\begin{table}[!h]
\centering
\small
\caption{Demographic distribution of $1.6$ million users in baseline dataset.}
\label{table:expected}
\begin{tabular}{|c|c|c|c|}
\hline
\multirow{2}{*}{\textbf{Race}} & \multicolumn{2}{c|}{\textbf{Gender}}    & \multirow{2}{*}{\textbf{Total}} \\ \cline{2-3}
                               & \textbf{Male}      & \textbf{Female}    &                                 \\ \hline
Asian                          & $120,950$ ($7\%$)  & $177,205$ ($11\%$) & $298,155$ ($18\%$)              \\ \hline
Black                          & $130,954$ ($8\%$)  & $107,827$ ($6\%$)  & $238,781$ ($14\%$)              \\ \hline
White                          & $538,625$ ($32\%$) & $595,302$ ($36\%$) & $1,133,927$ ($68\%$)            \\ \hline
\textbf{Total}                 & $790,529$ ($47\%$) & $880,334$ ($53\%$) & $1,670,863$ ($100\%$)           \\ \hline
\end{tabular}
\end{table}

\subsection{Gathering Social Connections and Interactions} \label{subsec:followers}

Ideally, to study how different demographic groups are connected and interact with each other, we would like to have at our disposal the followers (others who are following the user) and friends (others who are followed by the user) of all users from our dataset. However, the number of followers and friends of the $1.6$ million users in our dataset surpasses $6.4$ billion users, which is unfeasible to be crawled through the Face++ API. Next, we discuss a sampling procedure we used to create a dataset of users, their connections and their interactions with demographic information.  

For this specific analysis, we randomly selected a total of $6,000$ users from our dataset, $1,000$ users of each demographic group (i.e. White male, White female, Black male, Black female, Asian male, and Asian female). Then, we gathered their friends. As some of these users have a prohibitive large number friends, we limited our crawl to gather only the most recent $5,000$ friends of a given user, as this is the maximum number of user IDs the Twitter API returns per request. However, this strategy gathered the entire friends' list for $98.51\%$ of the users. 

Then, we follow the same methodology we discussed before to gather the demographic information of users. Firstly, we remove users that are not located in U.S., and then we attempt to identify the demographic aspects of each user using Face++ API. 
It is undeniable that the aggregated number of friends is extremely high and it is difficult to gather due to the limitation in term of requests that Face++ API allows us to make. Thus we were limited to gather demographic of at least $5\%$ of the friend's list.

Our social connections and interactions dataset contains $428,697$ users with the proper demographic information identified. Table~\ref{table:number_friends} presents the total number of the friends gathered for each demographic group that we examined. The average and median percentage of friends of the $6,000$ users for which we were able to gather demographic information are $10.15\%$ and $9.40\%$, respectively. We note that these fractions are usually higher than $5\%$ as some extra users were previously gathered in our $1.6$ million demographic dataset.

Also, we worked in a similar way for the analysis of interactions between the demographic groups. More specifically, we check the users that retweet and mention the tweets of the randomly selected users. To do that, we gathered all tweets (limited to a maximum of $3,200$ tweets due to Twitter REST API limitation\footnote{\url{https://dev.twitter.com/rest/reference/get/statuses/user\_timeline}}) of our $6,000$  users. We then identified the users who were mentioned or retweeted and we limited to gather the demographic information of only $5\%$ of retweeters and mentioned users for our analysis. Table~\ref{table:number_interactions} summarizes the amount of crawled users for each demographic group.

Our study about the connections among demographic groups (Section~\ref{sec:results_interconnections}) is based on this specific dataset.


\begin{table}[!ht]
 \centering
\caption{Number of Friends in each Group}
\label{table:number_friends}
\begin{tabular}{| c | c | c | c | c |}
\hline
 & \textbf {White} & \textbf{Black}& \textbf{Asian}& \textbf{Total} \\
\hline
\textbf{Male} &  $151,840$ & $52,437$ & $24,299$  & $228,576$  \\
\hline
\textbf{Female} & $137,010$  & $31,011$ & $32,100$  & $200,121$ \\
\hline
\textbf{Total} & $288,850$  & $83,448$  & $56,399$  & $428,697$ \\
\hline
\end{tabular}
\end{table}

\begin{table}[!ht]
 \centering
\caption{Number of Interactions in each Group}
\label{table:number_interactions}
\begin{tabular}{| c | c | c | c | c |}
\hline
 & \textbf {White} & \textbf{Black}& \textbf{Asian}& \textbf{Total} \\
\hline
\textbf{Male} & $246,879$  & $109,744$ & $51,370$ & $407,993$ \\
\hline
\textbf{Female} & $202,338$  & $60,108$ & $71,137$ & $333,583$ \\
\hline
\textbf{Total} & $449,217$ & $169,852$ & $122,507$ & $741,576$ \\
\hline
\end{tabular}
\end{table}

\subsection{Potential Limitations}

The gender and race inference are challenge tasks, and as other existing strategies have limitations and the accuracy of Face++ inferences is an obvious concern in our effort. The limitations of our data is discussed next.

\quad \quad \textbf{Accuracy of the inference by Face++:} Face++ itself returns the confidence levels for the inferred gender and race attributes, and it returns an error range for inferred age. In our data, the average confidence level reported by Face++ is $95.22\pm0.015\%$ for gender and $85.97\pm0.024\%$ for race, with a confidence interval of $95\%$. Recent efforts have used Face++ for similar tasks and reported high confidence in manual inspections~\cite{an2016greysanatomy,Zagheni2014,chakraborty2017@icwsm}. In addition, in recent scientific effort~\cite{chakraborty2017@icwsm} they evaluated the effectiveness of the inference made by Face++, using human annotators to label randomly selected profile images from Twitter. They measured the inter-annotator agreement in terms of the Fleiss’ $\kappa$ score which was $1.0$ and $0.865$ for gender and race, respectively.

Our dataset may also constitute the existence of fake accounts and bots. Previous studies provide evidence for an important rate of fake accounts \cite{Freitas2015@asonam,messias13@firstmonday} in Twitter.


\quad \quad \textbf{Data:} Finally, we note that our approach to identify users located in U.S. may bring together some users located in the same U.S. time zone, but from different countries. We, however, believe that these users might represent a small fraction of the users, given the predominance of active U.S. users in Twitter~\cite{sysomos}. Also, we are using the $1\%$ random sample of all tweets. However, the $1\%$ random sample is not the best data to capture all the dynamics happening in Twitter, its limitations are known~\cite{morstatter2014biased} and it is the best available option at our disposal.

\section{Inequalities in Visibility} \label{sec:results_inequalities}

We are interested to analyze the association of demographic status with visibility and discover possible inequalities. These asymmetries can be derived from the prejudices and stereotypes in the selection of which user will follow based on gender or race. Nilizadeh \textit{et al.}~\cite{nilizadeh2016twitter} provide evidence of inequalities and asymmetries in terms of visibility between males and females. 

We focus only in the visibility in the social network and not in user's influence of the audience which his/her is exposed. For our purpose we use two different features to measure a user visibility: its follower count and its listed count. In other words, we test whether the gender and race affect users' number of followers and how many lists they are added to.

The number of followers measures the real size of the audience that someone is exposed~\cite{icwsm10cha}. Finally, listed count represents the amount of times a user was added into a specific list by others in Twitter. Twitter Lists allow users to group and organize Twitter accounts that tweet on a topic that is of interest to her, and follow their collective tweets. Many users carefully create Lists to include other Twitter users who they consider as experts on a given topic. Previous research efforts~\cite{sharma2012inferring} attempt to gather Lists in large scale to find experts in Twitter.

As a baseline for comparisons, Table~\ref{table:expected} shows the demographic breakdown of users in our dataset across the different demographic groups. We can note a prevalence of females ($52.69\%$) in comparison to males ($47.31\%$) and a predominance of Whites ($67.86\%$) in comparison to Blacks ($14.29\%$) and Asians ($17.85\%$). This means that if we pick users randomly in our dataset, we would expect demographic groups with these proportions. We used these proportions next as a baseline for characterizing inequalities in Twitter.

\subsection{Gender Inequality}

\begin{figure}[!htb]
  \centering
    \includegraphics[width=0.4\textwidth]{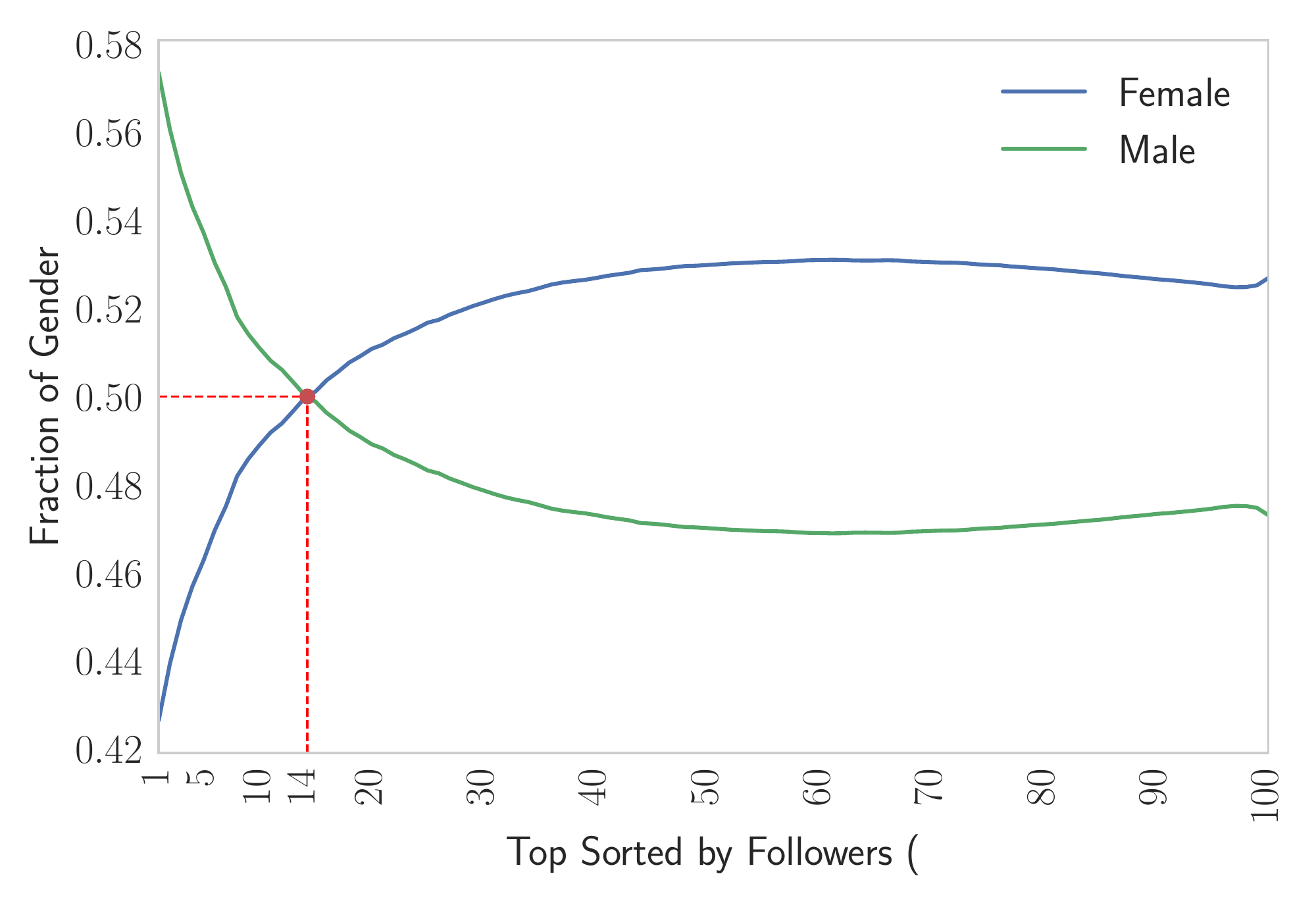}
  \caption{Distribution of the fraction of males and females in the rank of users with most followers. The intersection point happens in the Top $14\%$, meaning the fraction of females became higher than males after this point.}
  \label{fig:top_gender}
\end{figure}

We begin our analysis by sorting the number of followers of all users in our dataset and we check the fraction of males and females in an incremental proportion of top followed users. Figure~\ref{fig:top_gender} provides some insight on this point. Although, the number of females is higher than the males as shown in Table~\ref{table:expected}, males tend to be among the top followed users. The most significant difference exists in top $1\%$ with a fraction of $57.34\%$ of males and $42.65\%$ of females, which is almost $15\%$ and this difference is decreasing until the top $14\%$, where the fraction of females become higher than the fraction of males.

\begin{figure}[!htb]
  \centering
    \includegraphics[width=0.47\textwidth]{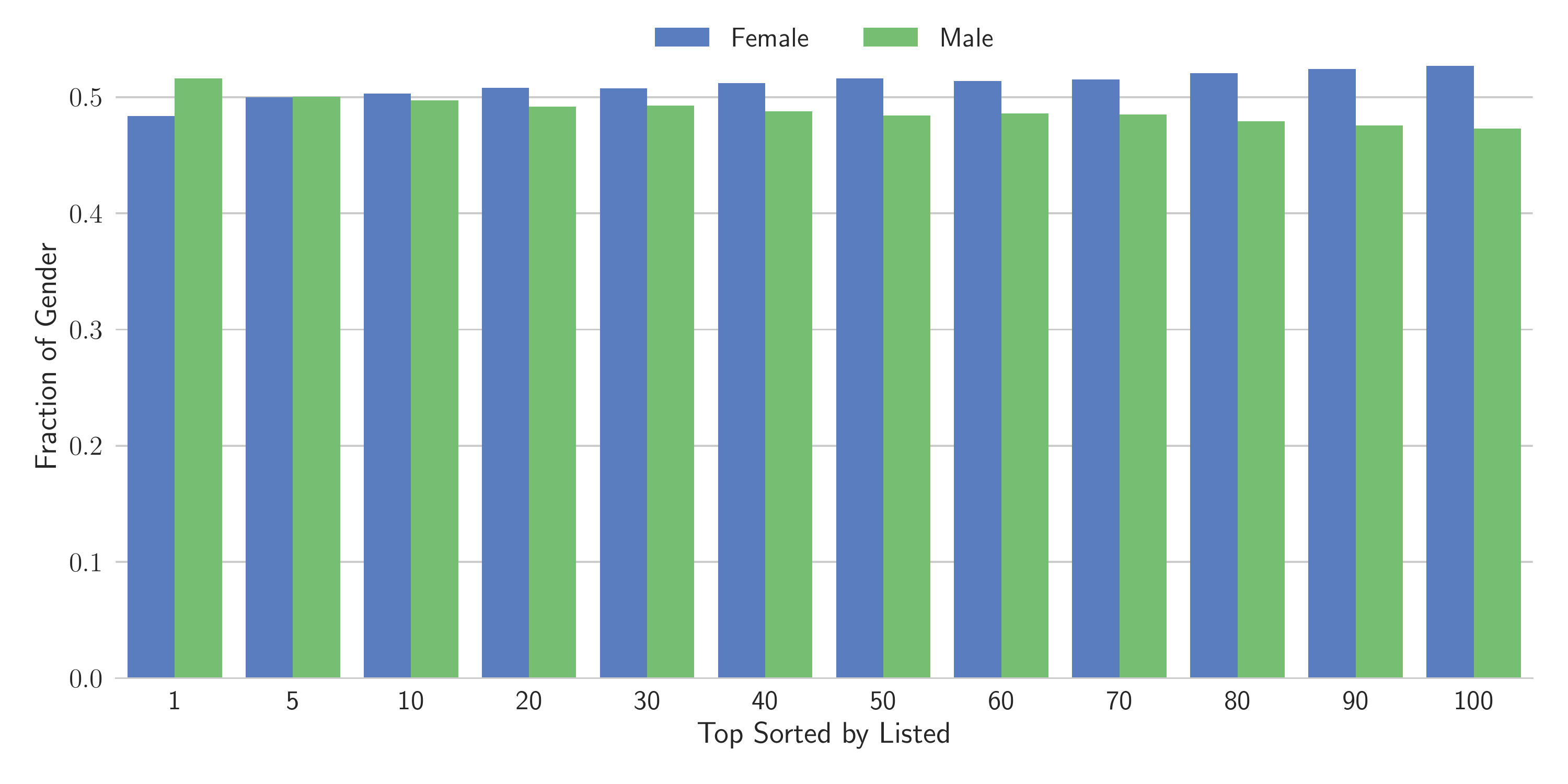}
  \caption{Distribution of the fraction of males and females in the rank of most listed users.}
  \label{fig:top_gender_list}
\end{figure}

The same kind of observation can be made for the rank of lists, as shown in Figure~\ref{fig:top_gender_list}. However, the discrepancy perceived is smaller in the lists basically restricted to the top $1\%$ most listed users with $51.23\%$ of males and $48.64\%$ of females. In top $6\%$ the fraction of females become higher than males. The same occurs in top $8\%$. The results of this analysis reflect the general idea that higher positions are usually taken by males \cite{sugimoto2013global}. 

Our observations also reinforce previous findings related to gender in Twitter~\cite{nilizadeh2016twitter}, suggesting the existing of the so called 'glass ceiling effect', which posits that females face an invisible barrier at the highest levels of an organization~\cite{cotter2001glass}. One concern raised by authors on that work was that disparity may be driven by a small number of 'elite' users. Our findings diminish such concerns as our dataset is at least two order of magnitudes larger and we noted such discrepancies even in the top $10\%$ positions (i.e. top $167.086$ users). 





\subsection{Race Inequalities}

\begin{figure}[!htb]
  \centering
    \includegraphics[width=0.47\textwidth]{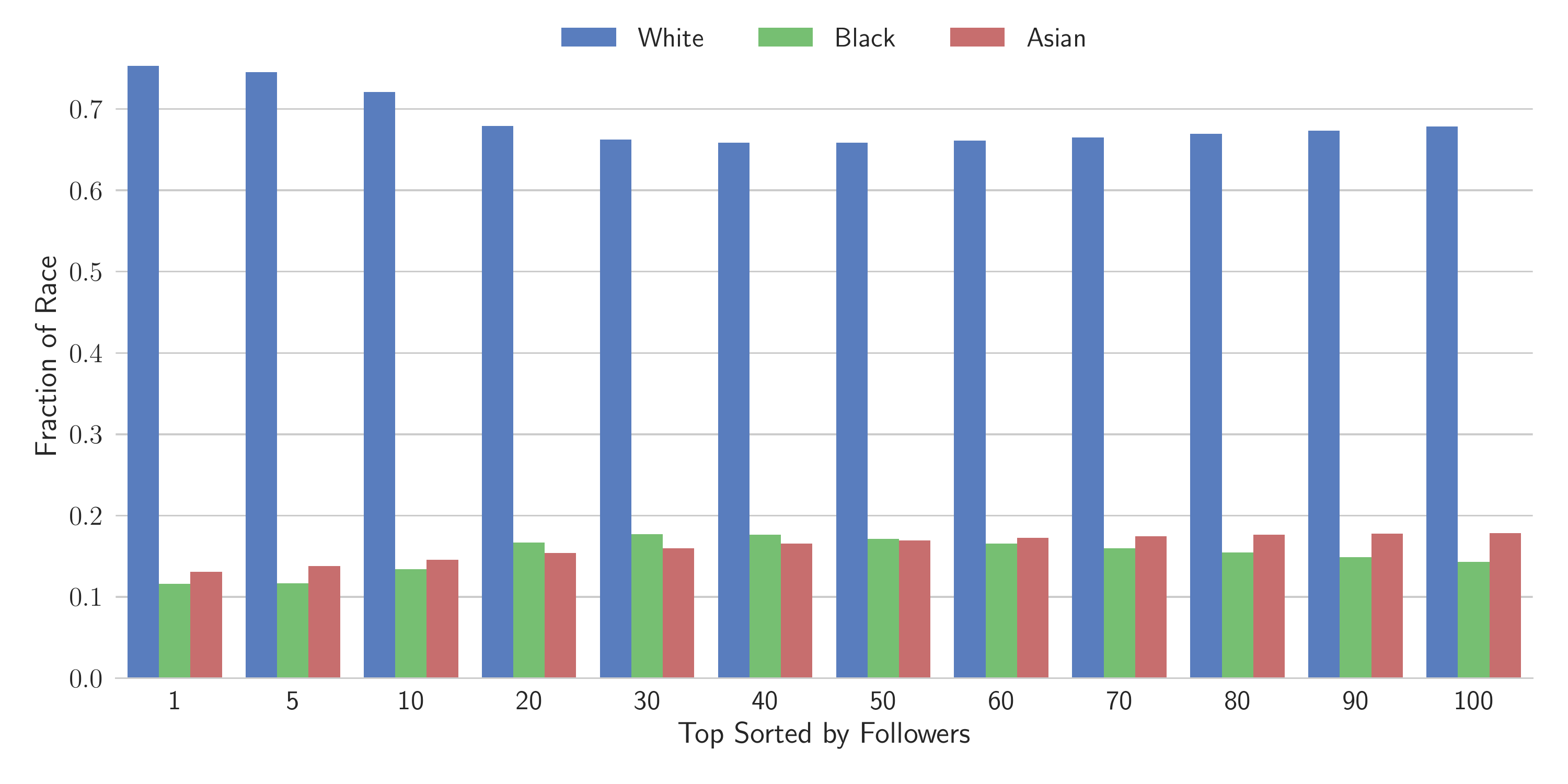}
  \caption{Amount of Whites, Blacks, and Asians in the rank of users with most followers}
  \label{fig:top_race}
\end{figure}

\begin{figure}[!htb]
  \centering
    \includegraphics[width=0.47\textwidth]{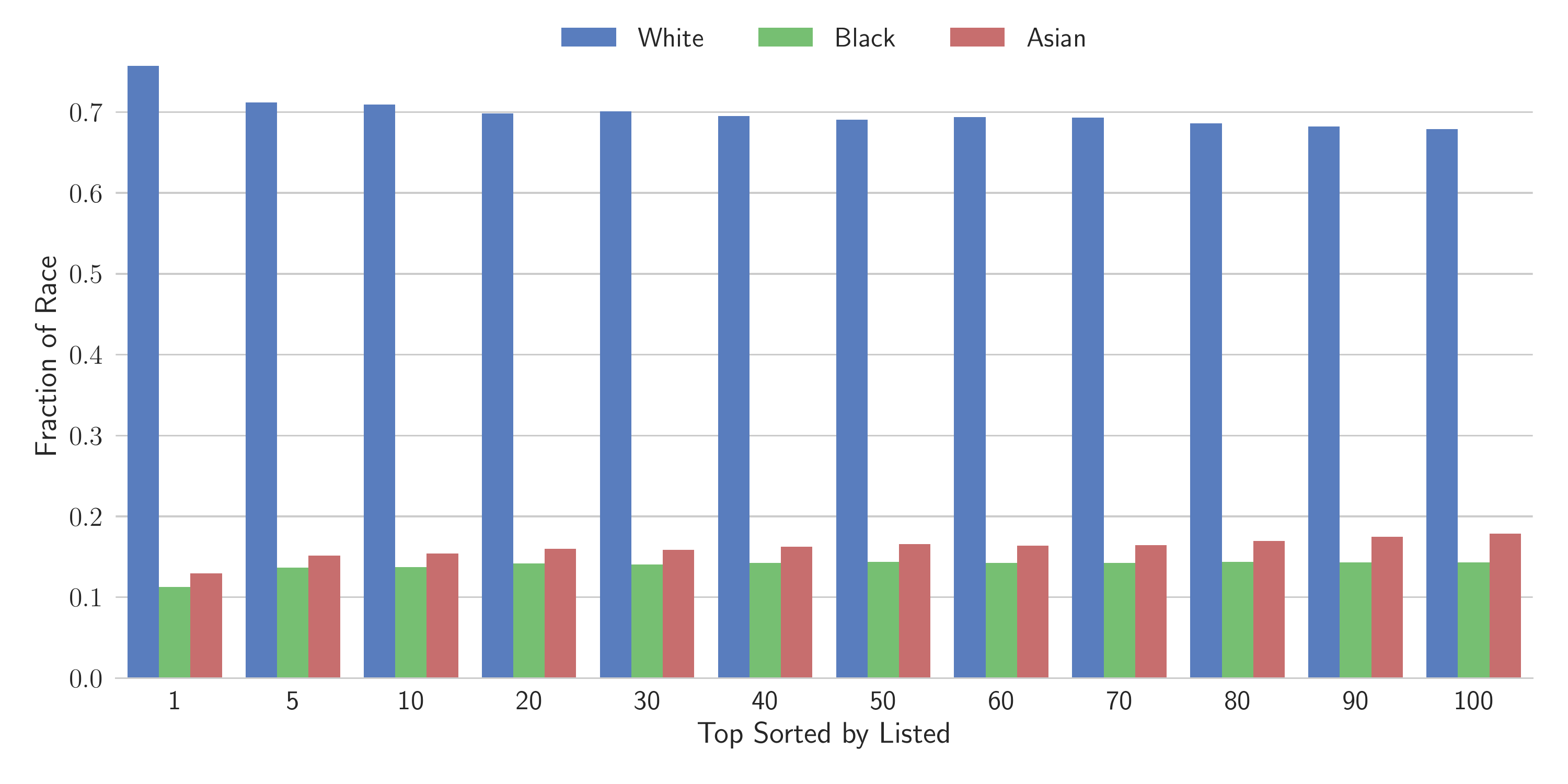}
  \caption{Amount of Whites, Blacks, and Asians in the rank of the most listed users}
  \label{fig:top_race_list}
\end{figure}

We now turn our attention to the analysis of race inequalities. 
Similarly to what we have done in the previous section, we examine here the presence of each user race within the top positions in a rank of top followed users and top listed users. Figure~\ref{fig:top_race} and Figure~\ref{fig:top_race_list} provide the necessary information about race and top followed users. We can note that the amount of White users tend to be higher in the both top positions. In the rank of users with more followers and in the rank of most listed users.

Our results suggest that the race disparity in Twitter visibility occurs, meaning that at the highest levels of visibility, users perceived to be White come out on top position. This observation reinforces many previous observations related to race inequality in United States~\cite{bonilla2006racism,oliver2006black}. 





\subsection{Taking Together Gender and Race Inequalities}

\begin{figure}[!htb]
  \centering
    \includegraphics[width=0.47\textwidth]{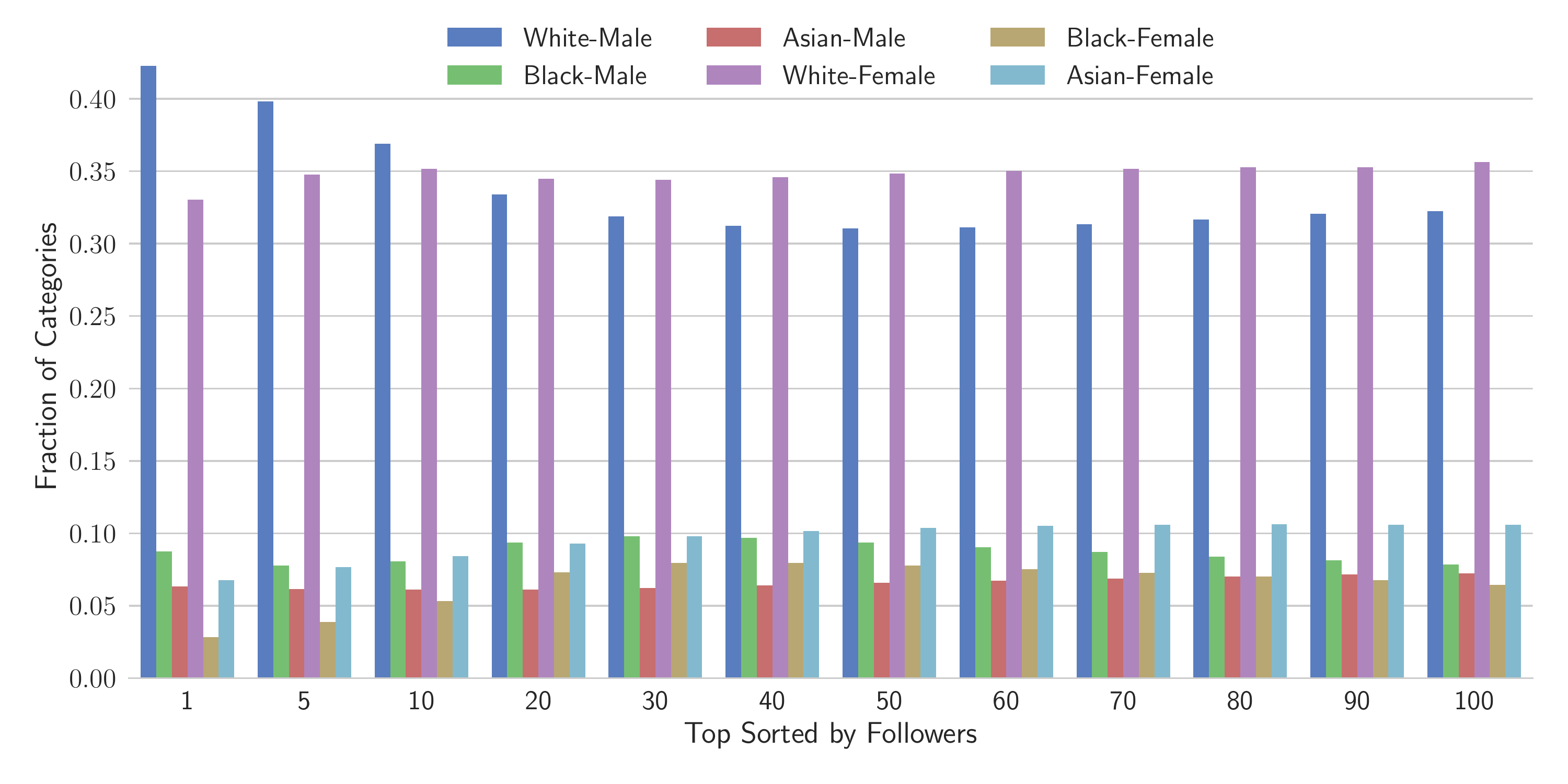}
  \caption{Demographic groups in terms of gender and race as a function of the rank of the most listed users}
  \label{fig:top_categories_followers}
\end{figure}

\begin{figure}[!htb]
  \centering
    \includegraphics[width=0.47\textwidth]{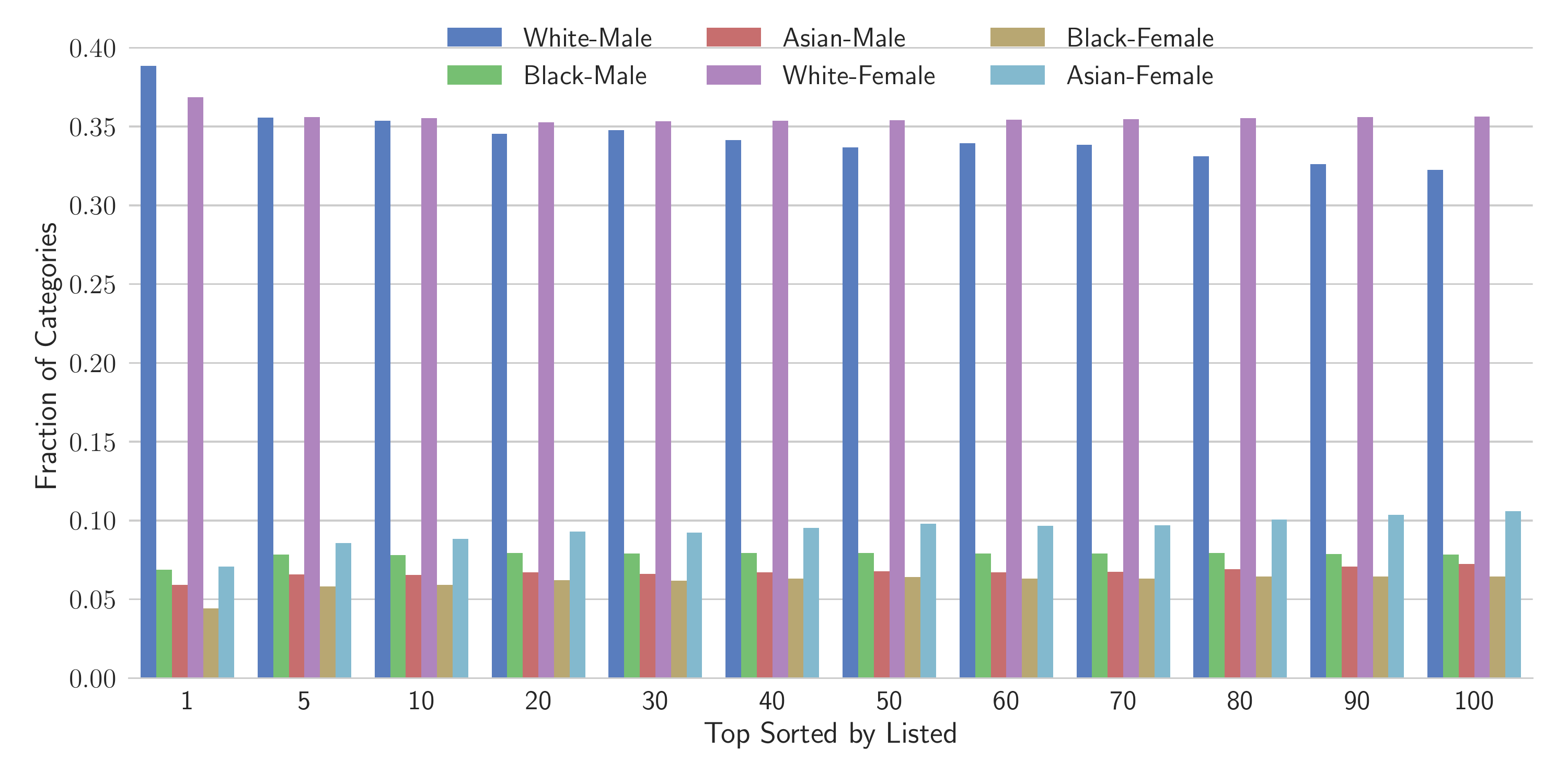}
  \caption{Demographic groups in terms of gender and race as a function of the rank of the most listed users}
  \label{fig:top_categories_listed}
\end{figure}

Finally, we attempt to quantify the existing inequality when we look into gender and race aspects at the same time. We note from Figure~\ref{fig:top_categories_followers} that the amount of White males among the top $1\%$ users with more followers is $42.27\%$. This represents an increase of $10.04\%$ in the fraction of White males in comparison to the expected proportion of White males in our dataset, which is $32.23\%$. We can note that populations of Black females and Asian females experience a reduction of $3.60\%$ and $3.84\%$ in the top 1\% most followed users, respectively. We note similar trends in terms of lists as we can see Figure~\ref{fig:top_categories_listed}.

To better analyze to what extent demographic groups in the top $1\%$ is substantially higher or lower than what one would expect based on our demographic information dataset, Table~\ref{table:top_followers_listed_diff} shows the increase or decrease in the proportion of each demographic group in comparison with such baseline expected population. Overall, this results quantify the perceived inequalities that six demographic groups experience  
at the highest levels of visibility (i.e. top $1\%$ most followed and listed). The most privileged demographic group in the rank of followers is the group of White males, with more than $20\%$ of users than a baseline expected random sample. The amount of White females is also higher than the baseline, but much smaller amount (only $3\%$) in comparison to males. The most unprivileged groups are Asian females and Black females, underrepresented in the top $1\%$ by more than $31\%$.  

\begin{table}[!h]
\centering
\caption{Relative proportion of each demographic group in the top $1\%$ rank of users with more followers and the most listed in comparison to the baseline expected population.}
\label{table:top_followers_listed_diff}
\begin{tabular}{|l|c|c|c|c|}
\hline
\multirow{2}{*}{\textbf{Race}} & \multicolumn{2}{c|}{\textbf{Followers (\%)}}                    & \multicolumn{2}{c|}{\textbf{Listed (\%)}}                       \\ \cline{2-5} 
                               & Male                          & Female                       & Male                          & Female                       \\ \hline
Asian                          & \textcolor{red}{-12.70}  & \textcolor{red}{-36.25} & \textcolor{red}{-18.33}  & \textcolor{red}{-33.32} \\ \hline
Black                          & \textcolor{blue}{+11.68} & \textcolor{red}{-55.83} & \textcolor{red}{-12.36}  & \textcolor{red}{-31.24} \\ \hline
White                          & \textcolor{blue}{+31.16} & \textcolor{red}{-7.27}  & \textcolor{blue}{+20.53} & \textcolor{blue}{+3.45} \\ \hline
\end{tabular}
\end{table}

\section{Demographic Group Interconnections} \label{sec:results_interconnections}

Next, we analyze how demographic groups are connected with each other. Our ultimate goal is to examine the proportion of connections and interactions among each demographic group. Thus, for this analysis we used the dataset related to friends, followers and interactions discussed previously.

In order to represent connection probabilities among demographic groups we built a probabilistic graph and we compute the probability of a connection/interaction for every pair of users.
We declare as probability the empirical probability of the existence of the specific  event in our dataset.

To gain a holistic view, we aggregate all pairs of users making the nodes to represent demographic groups and each direct edge to represent the probability of a relationship happened, like friendship or interaction. The sum of all outgoing probabilities for each demographic group is $1.0$. 

\subsection{Gender and its Interconnections}

We begin by analyzing interconnection among groups of users separated by Gender. 
We analyze the probability of males and females users to be friends with (i.e. to follow) other males and females users. In our analysis, we can note that females tend to equally follow males and females in $50\%$, while males tend to follow $56\%$ of males against $44\%$ of females. If we take the numbers in Table~\ref{table:expected} as a proxy for the expected population of males and females in a randomly created list of friends, we note that even the equally distributed list of friends of females is actually disproportional as the expected percentage of males and females are $47.31\%$ and $52.69\%$, respectively. This means that males tend to follow $18.37\%$ more other males than expected, whereas females tend to follow $5.7\%$ more males than expected. This suggests that both, males and females connections, take their part of responsibility on these gender inequalities, but it also shows that males tend to take the larger part of the responsibility. 

In terms of interactions, similar observations can be made. However, the proportion of males and females retweet/mentioned by females is quite close to the expected population ($53\%$ of females against $47\%$ for males). For the group of males, we noted a much higher disparity in terms of interactions in comparison to friendship. Males tend to interact with $64\%$ of other males and only $36\%$ of females. 

This result quantifies the perceived inequality in Twitter. These observations are inline with perceived inequalities in the offline world, including in academic job hiring, earnings, funding and academic rating~\cite{sugimoto2013global,sugimoto2015relationship} as well as income differences between males and females in the population of high earners \cite{merluzzi2015unequal}.

\subsection{Race and its Interconnections}

\begin{figure}[!htb]%
    \centering
    \subfloat[Friendship]{{\includegraphics[width=4.1cm]{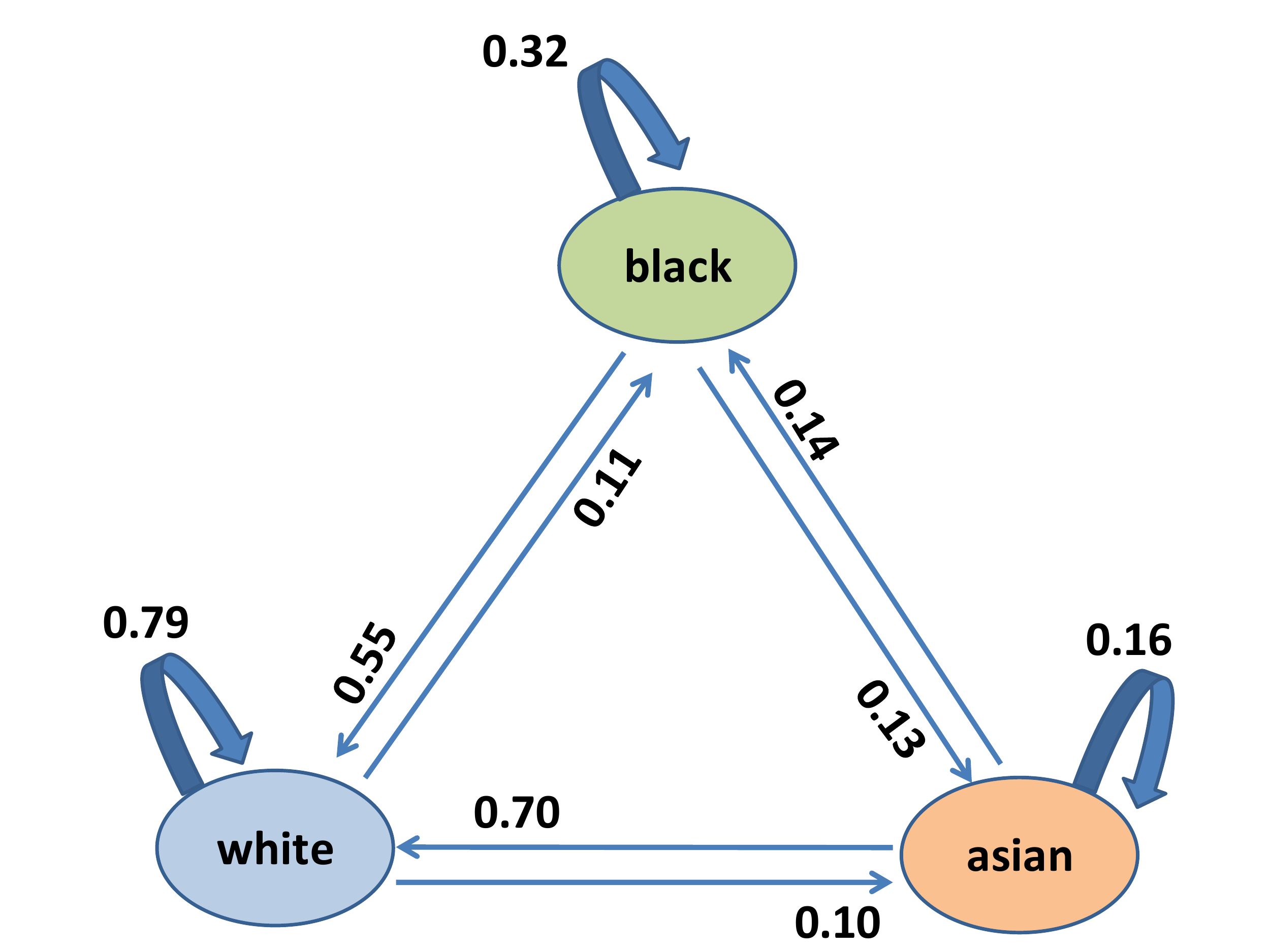} }}%
    \subfloat[Interaction]{{\includegraphics[width=4.1cm]{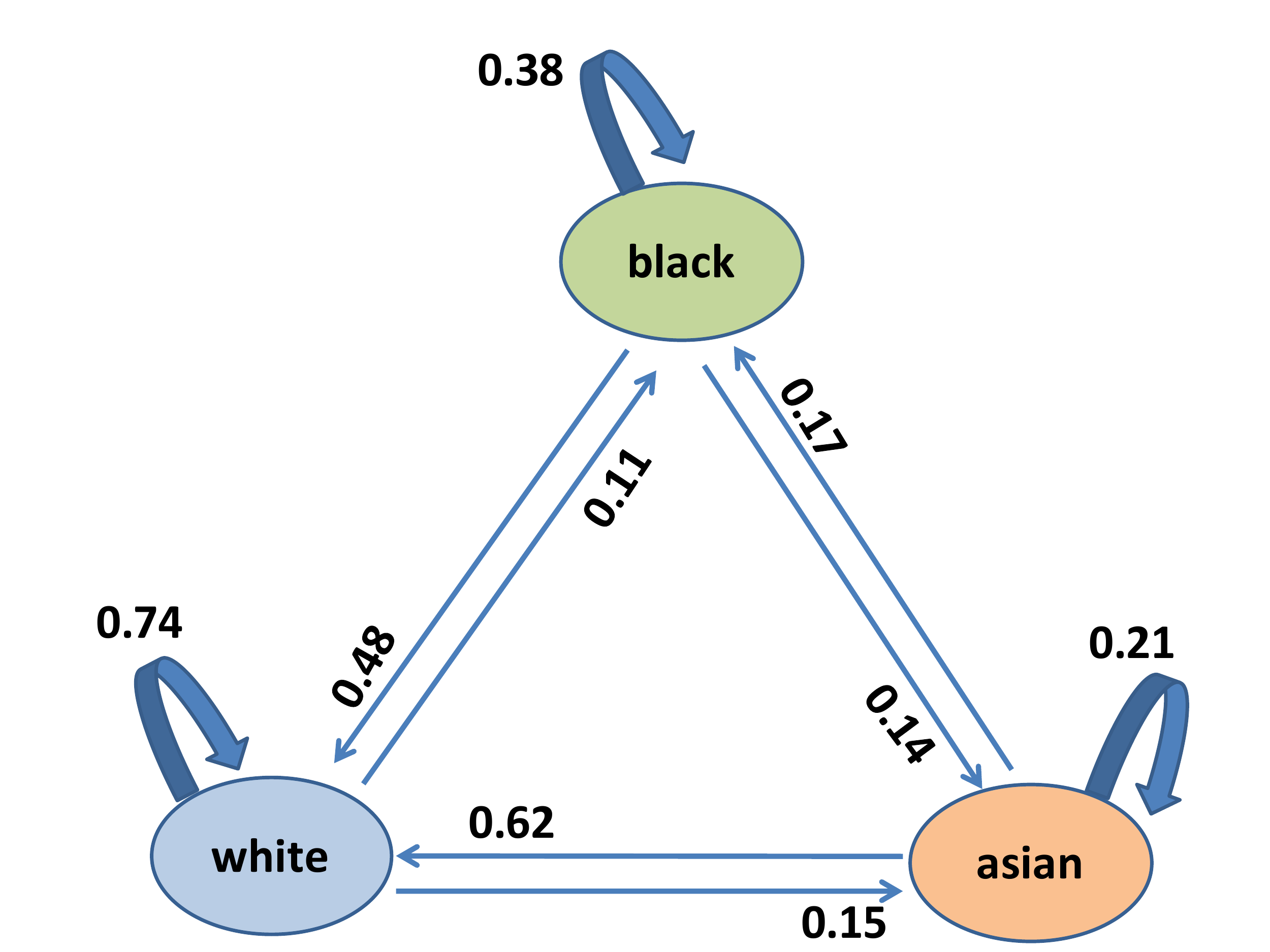} }}%
    \caption{Probability of friendship and interaction for demographic groups of users separated by race}%
    \label{fig:race_graph_friendship_inter}%
\end{figure}

Next, we turn our attention to the analysis of interconnection among groups of users separated by race. Figure~\ref{fig:race_graph_friendship_inter} shows these interconnections in terms of friendship and interactions, respectively. 

We start by analyzing the self loops on each node.  We note that White users tend to follow about $79\%$ of other White users, Black users tend to follow $32\%$ of Black users and Asians tend to follow $16\%$ of Asians. This means that the key to analyze these Figures is to recall the reference distribution of users in our dataset according to their race, in which Whites represent $67.86\%$, Black accounts to $14.29\%$, and Asian accounts to $17.85\%$. Thus, we can note that, in comparison to the expected distribution of users, Whites tend to follow $16.41\%$ more Whites than expect, Blacks tend to follow impressive $2.24$ times more  Blacks than expect, and Asians end up following less ($10.26\%$) Asians than expect. In other words, the expected homophily was not clear for the case of Asians.

Similar observations can be made for interactions, only exception for among Asians, where the fraction of interactions is higher than the reference distribution. Although Asians tend to follow more White users (even more than the reference distribution), the interactions with White users are below the reference fraction. Overall, Whites and Blacks showed quite high numbers of endogenous connections and interactions as both groups seem to avoid following a different race. The stronger mutual interconnection is between Black users. 

\subsection{Demography of Interconnections}

\begin{figure}[!htb]
  \centering
    \includegraphics[width=0.40\textwidth]{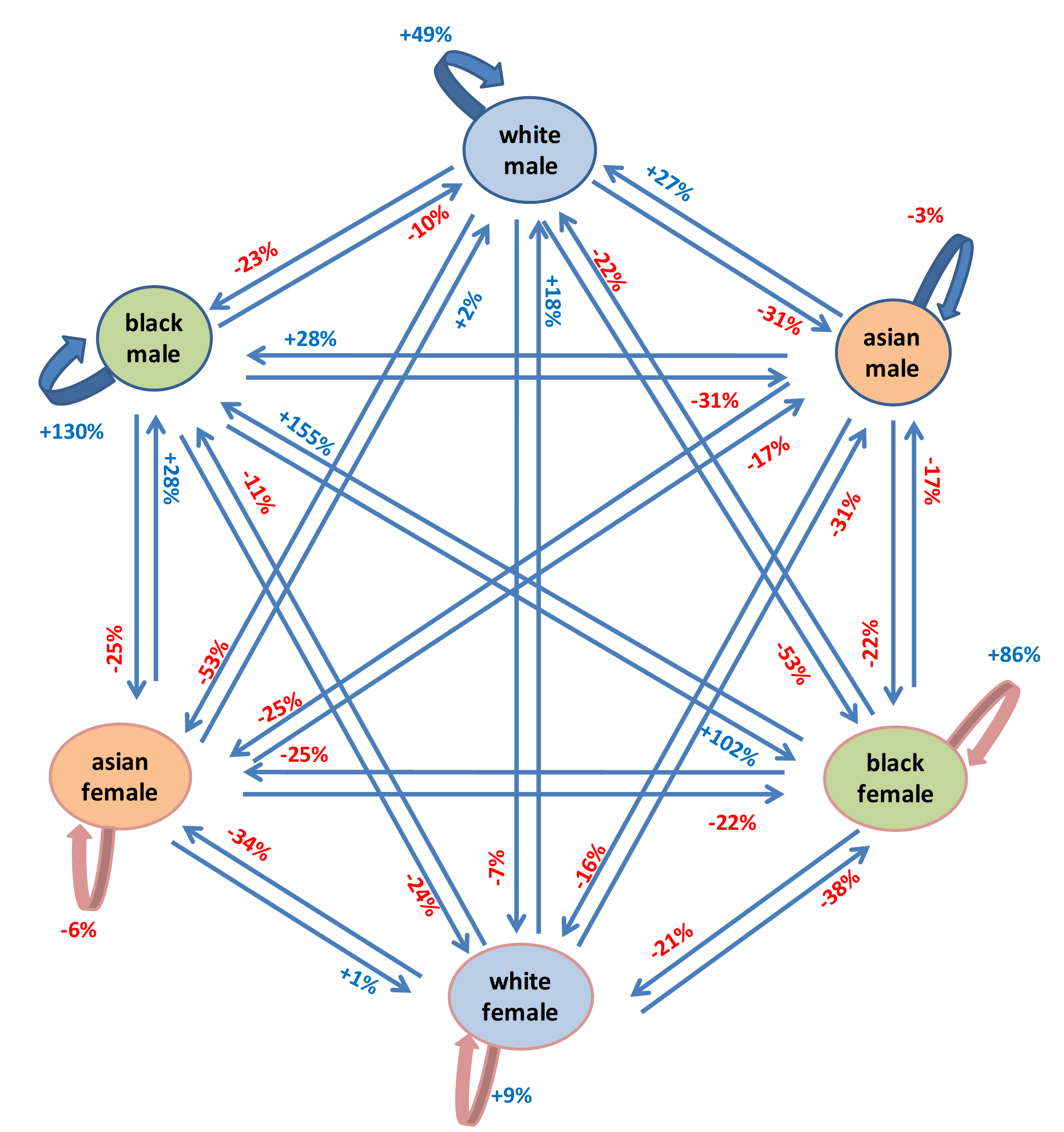}
  \caption{Relative probability of friendship for groups of users separated by race and gender. The value declares the relative increase (\textcolor{blue}{blue} color) or decrease (\textcolor{red}{red} color) in comparison to the expected demographic population.}
  \label{fig:all_graph_friends}
\end{figure}

\begin{figure}[!htb]
  \centering
    \includegraphics[width=0.40\textwidth]{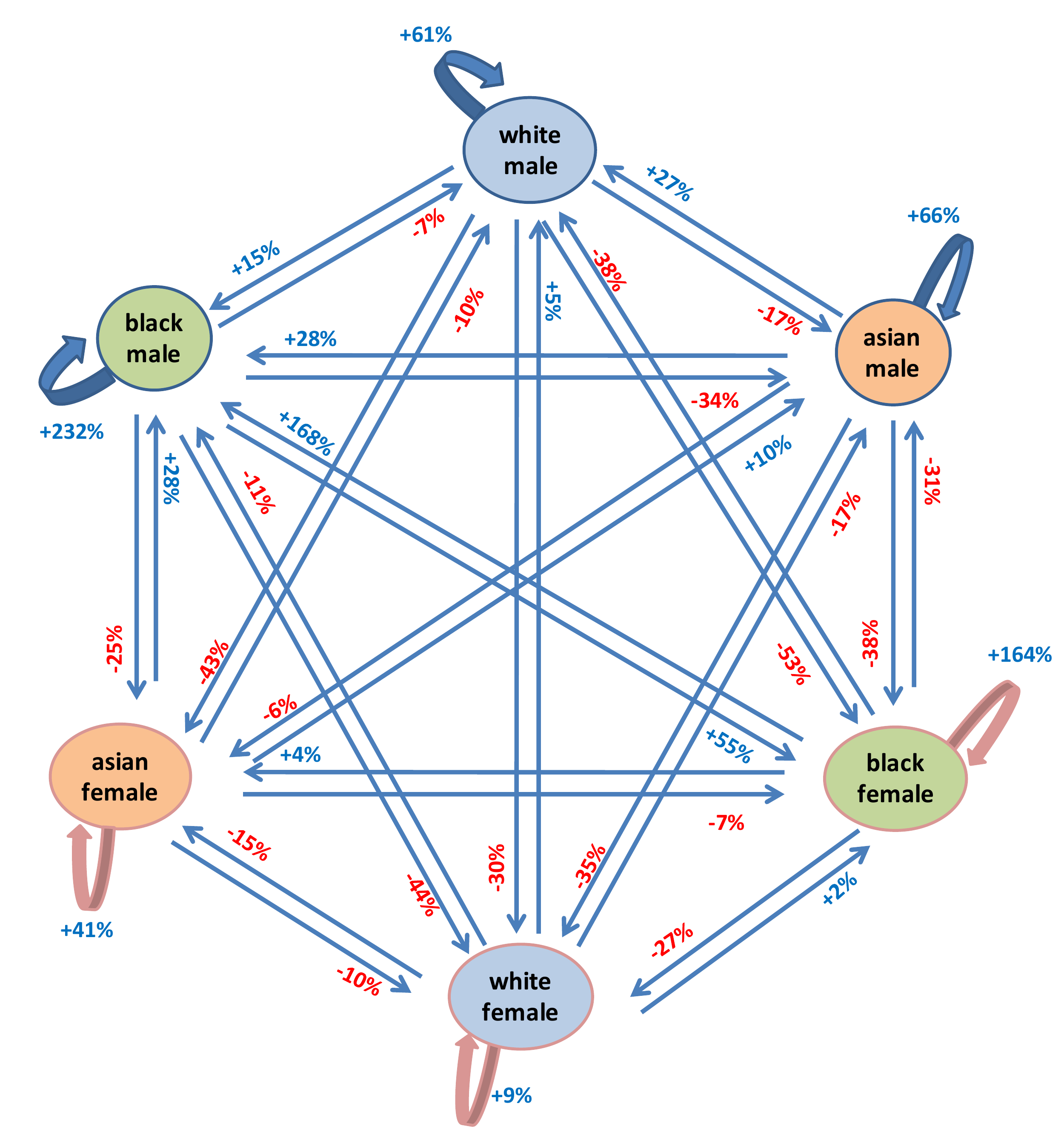}
  \caption{Relative probability of interactions for groups of users separated by race and gender. The value declares the relative increase (\textcolor{blue}{blue} color) or decrease (\textcolor{red}{red} color) in comparison to the expected demographic population.}
  \label{fig:all_graph_act}
\end{figure}

Finally, we characterize the interconnection among groups of users both separated by race and gender. Figures~\ref{fig:all_graph_friends} and~\ref{fig:all_graph_act} show these interconnections in terms of friendship and interactions, respectively. However, given the high number of edges in these figures, instead of providing the transition probabilities, we computed their relative increase or decrease in comparison to the expected demographic population.

In addition to reinforce previous findings from the analysis of gender and race separately presented before, these results show interesting trends. First, we note that White males tend to receive only positive edges from White and Asian groups, meaning that only Black users (both males and females) tend to have a fraction of White males users as friends that are relatively low than the expected demographic distribution in Twitter. This suggests that race plays a more important role for connections with White males. In terms of interactions, similar observations hold, except that Asian females tend to have a relatively smaller number of interactions with White males. Interestingly, White females appear over-represented only in the list of friends of Asia females.  

Another interesting pattern is that the highest values for self-loops are for Black males ($130\%$) and Black females ($86\%$), which is even higher for interactions, $232\%$ and $164\%$, respectively. The interconnections between these two groups are also impressively large. On the other hand, the lower values for self-loops are for Asians, with negative values for friendship, but positive for interactions, as discussed before. Interestingly, all incoming links to the two Asian groups are negative, meaning that the other two demographic groups tend to have an under-represented proportion of Asians as friends and in terms of interactions. 

\section{Concluding Discussion} \label{sec:conclusion}

This paper characterizes demographic information of users in Twitter, identifying inequalities and asymmetries in gender and race. To do that we gather gender and race of Twitter users located in U.S. using advanced image processing algorithms from Face++. We collect a large sample of $1,670,863$ users located in U.S. with identified demographic information. We have also crawled a large sample of tweets and friends of these to study connections and interactions among demographic groups of users. 

Our analysis reinforces evidence about gender inequality in terms of visibility and introduces race as a significant demographic factor, which reveals hiding prejudices between groups of different demographic status. More important, our effort is the first of a kind to quantify to what extent one demographic group follows and interact with each other and the extent to which these connections and interactions reflect in inequalities in Twitter. 

Among our findings, we show that the Twitter glass ceiling effect, typically applied to females, also occurs in Twitter for males, if they are Black or Asians. Our analysis of connections and interactions among these groups explain part of causes of such inequalities and offer hints for the promotion of equality in the online space. 

There are some future directions we would like to pursue next. First, we plan to explore age as a demographic factor in social media. Second, we believe that the increasing availability of information about demographic would help the development of systems that promote more diversity and less inequality to users.



\section*{Acknowledgments} \label{sec:acknowledgments}



\footnotesize{This research is funded by projects InWeb (grant MCT/CNPq 573871/2008-6) and MASWeb (grant FAPEMIG/PRONEX APQ-01400-14), by grants from CAPES, CNPq, FAPEMIG, and Humboldt Foundation. Also, this research was supported by the European Commission in the framework of Erasmus Mundus mobillity programme IBRASIL.}

\small
\bibliographystyle{SIGCHI-Reference-Format}
\balance
\bibliography{references}

\end{document}